\documentclass[12pt]{article}

\newcommand{\Equ}[1]{\begin{equation} \label{eq:#1}}
\newcommand{\EndEqu}{\end{equation}}

\newcommand{\spc}{\;\;\;\;\;\;\;\;\;\;}

\newcommand{\1}{\mbox{\rm 1 \hspace{-1.05 em} 1}}

\title{On the Construction of Zero Energy States in Supersymmetric 
Matrix Models\\II}
\author{Jens Hoppe\thanks{Heisenberg Fellow}\\
Institute for Theoretical Physics\\
ETH H{\"o}nggerberg\\
CH-8093 Z{\"u}rich}
\date{September/October 1997}

\begin{document}
\maketitle

\begin{abstract}
	For the $SU(N)$ invariant supersymmetric matrix model related to 
	membranes in $11$ space-time dimensions, the general (bosonic) 
	solution to the equations $Q_\beta^\dagger \Psi =0$ ($Q_\beta \Psi
=0$) is 
	determined.
\end{abstract}
\newpage
Continuing \cite{H}, I present the explicit form of all 
$SU(N)$-invariant wave functions
\begin{equation}
	\Psi \;=\; \psi \:+\: \frac{1}{2} \: \lambda_a \lambda_b \: 
	\psi_{ab} \:+\: \cdots \:+\: \frac{\lambda_{a_1} \cdots 
	\lambda_{a_\Lambda}}{\Lambda!} \: \psi_{a_1 \cdots a_\Lambda} \spc
,
	\label{1}
\end{equation}
$\Lambda \;=\; 8(N^2-1)$, $\{\lambda_a, \lambda_b \}=0=
\{\partial_{\lambda_a}, \partial_{\lambda_b} \}$, $\{\lambda_a, 
\partial_{\lambda_b} \}=\delta_{ab}$, satisfying $Q_\beta^\dagger \Psi =0$
($Q_\beta \Psi =0$), $\beta=1,\ldots,8$ arbitrary (but fixed), when
\begin{equation}
	Q_{\beta}\;=\; M_a^{(\beta)} \: \lambda_a \:+\: D_a^{(\beta)} \: 
	\partial_{\lambda_a} \;\;,\;\;\;
	Q_{\beta}^\dagger\;=\; M_a^{(\beta) \:\dagger} \:
\partial_{\lambda_a} \:+\:
	D_a^{(\beta)\:\dagger} \: \lambda_a
	\label{2}
\end{equation}
with
\begin{eqnarray}
	M_{\alpha A}^{(\beta)} & = & \delta_{\alpha \beta} \: i q_A \:+\:
i 
	\Gamma^j_{\alpha \beta} \: \frac{\partial}{\partial x_{j A}} \:-\: 
	\frac{1}{2} \: f_{ABC} \:x_{jB}\:x_{kC}\:\Gamma^{jk}_{\alpha
\beta}
	\label{3}  \\
	D_{\alpha A}^{(\beta)} & = & \delta_{\alpha \beta} \: 2 \partial_A 
	\:-\: i f_{ABC} \:x_{jB} \: \overline{z_C} \Gamma^j_{\alpha \beta}
	\spc ,
	\label{3a}
\end{eqnarray}
$\alpha, \beta=1,\ldots,8$, $j,k,l=1,\ldots,7$, $A,B,C=1,\ldots,N^2-1$,
$\{\Gamma^j, \Gamma^k\}=2 \delta^{jk} \1$, 
$\overline{\Gamma^j}=-\Gamma^j$.

In order to resolve the non-commutativity of $M$ with $M^\dagger$,
consider
\begin{equation}
	\tilde{Q}_\beta^\dagger \;:=\; F_\beta^{-1} \:Q_\beta^\dagger \:
F_\beta
	\label{4}
\end{equation}
with
\begin{equation}
	F_\beta \;=\; \exp \left( \frac{1}{6} \:f_{ABC} \:x_{jA} \:x_{kB} 
	\:x_{lC} \: u_{jkl}(\beta) \right) \spc .
	\label{5}
\end{equation}
While (\ref{5}) commutes with $D^\dagger$, the unwanted 3rd term in
\begin{equation}
	M^{(\beta) \:\dagger}_{\alpha A} \;=\; -i \delta_{\alpha \beta}
\:q_A \:-\: 
	i \Gamma^j_{\alpha \beta} \: \frac{\partial}{\partial x_{jA}}
\:-\: 
	\frac{1}{2} \:f_{ABC} \:x_{jB} \:x_{kC} \:\Gamma^{jk}_{\alpha
\beta}
	\label{6}
\end{equation}
will be removed, provided
\begin{equation}
	\Gamma^j_{\alpha \beta} \:u_{jkl}(\beta) \;=\; i 
	\Gamma^{kl}_{\alpha \beta}
	\label{7}
\end{equation}
for all $\alpha, k, l$ (and arbitrary, but fixed $\beta$).

In the representation
\begin{equation}
	i \hat{\Gamma}^j_{k8}\;:=\; \delta_{jk}\;\;,\;\;\; i 
	\hat{\Gamma}^j_{kl} \;:=\; -c_{jkl} \spc ,
	\label{8}
\end{equation}
where the $c_{jkl}$ (totally antisymmetric) are octonionic structure 
constants satisfying
\begin{equation}
	c_{jkl} \:c_{mnl} \;=\; \delta^j_m \:\delta^k_n \:-\: \delta^j_n 
	\:\delta^k_m \:-\: \frac{1}{6} \:\epsilon_{jkmnrst} \: c_{rst}
	\label{9}
\end{equation}
(cp. \cite{W}, where this representation was already used in a 
truncation of the 11-dimensional model to $N=1$ sypersymmetry, with a 
non-normalizable zero-energy state of the form (\ref{5}))
the solution of (\ref{7}) turns out to be
\begin{equation}
	\hat{u}_{jkl}(\beta) \;=\; c_{jkl} \; \left\{ \begin{array}{cc}
	+1 & {\mbox{if $\beta=j,k,l$ or $8$}} \\
	-1 & {\mbox{if $\beta \neq j,k,l$ and $8$}} \end{array} \right.
	\label{10}
\end{equation}
(which is easy to check, using $\hat{\Gamma}^{jk}_{l8}=-c_{jkl}$ and
$\hat{\Gamma}^{kl}_{mn}=\delta^k_m \delta^l_n - \delta^k_n 
\delta^l_m +\frac{1}{6} \epsilon_{klmnrst} c_{rst}$; the 
representation invariant form of (\ref{10}) is $i (\Gamma^{jkl})_{\beta 
\beta}$).

In order to solve the equation $Q^\dagger_\beta \Psi = 0$ ($Q_\beta \Psi =
0$),
consider $\tilde{Q}^\dagger_\beta (F_\beta^{-1} \Psi)=0$, i.e.
\begin{equation}
	\left( \left(-i \delta_{\alpha \beta} \:q_A \:-\: i
\Gamma^j_{\alpha 
	\beta} \:\frac{\partial}{\partial x_{jA}} \right) 
	\frac{\partial}{\partial \lambda_{\alpha A}} \:+\: 
	D^{(\beta)\:\dagger}_{\alpha A} \:\lambda_{\alpha A} \right)
\tilde{\psi} 
	\;=\; 0 \;\;\;,
	\label{11}
\end{equation}
or, in components,
\begin{equation}
	(2k-1) \:D^{(\beta)\:\dagger}_{[a_1} \: \tilde{\psi}_{a_2 \cdots 
	a_{2k-1}]} \;=\; N^{(\beta) \:\dagger}_{a_{2k}} \:
\tilde{\psi}_{a_1 \cdots 
	a_{2k}} \spc ,
	\label{12}
\end{equation}
$k=1,2,\ldots K:=4(N^2-1)$ with
\begin{equation}
	N^{(\beta) \:\dagger}_{\alpha A} \;:=\; -i \delta_{\alpha \beta}
\:q_A 
	\:-\: i \Gamma^j_{\alpha \beta} \:\frac{\partial}{\partial x_{jA}} 
	\;=\; F_\beta^{-1} \:M^{(\beta)\:\dagger}_{\alpha A} \:F_\beta
	\;=\; -N_{\alpha A}^{(\beta)} \;\;\;.
	\label{13}
\end{equation}
Using
\begin{equation}
	J_E \: \tilde{\psi} \;:=\; -i f_{EAA^\prime} \left( x_{jA} 
	\:\frac{\partial}{\partial x_{jA^\prime}} \:+\: z_A \:
	\partial_{A^\prime} \:+\: \overline{z_A} \: 
	\overline{\partial_{A^\prime}} \:+\: \lambda_{\alpha A}\: 
	\partial_{\lambda_{\alpha A^\prime}} \right) \tilde{\psi} \;=\; 0
	\label{14}
\end{equation}
one can show that the \underline{general} solution of
(\ref{11})/(\ref{12}) is
\begin{equation}
	\tilde{\psi}_{a_1 \cdots a_{2k}} \;=\; -(2k) (2k-1) 
	\:N_{[a_1}^{(\beta)} \:(N^+ N)^{-1} \: D_{a_2}^{(\beta)\:\dagger} 
	\tilde{\psi}_{a_3 \cdots a_{2k}]} \:+\: \tilde{\psi}^{(h)}_{a_1
\cdots 
	a_{2k}} \;\;\;,
	\label{15}
\end{equation}
hence the general solution of $Q^\dagger_\beta \Psi=0$
\begin{equation}
	\psi_{a_1 \cdots a_{2k}} \;=\; -(2k)(2k-1) \:F_\beta \:N_{[a_1}\:
	(N^\dagger N)^{-1} \: D_{a_2}^{\dagger} \:F_\beta^{-1} \:\psi_{a_3
\cdots 
	a_{2k}]} \:+\: \psi^{(h)}_{a_1 \cdots a_{2k}} \;\;\;,
	\label{16}
\end{equation}
where $N^{(\beta)\:\dagger}_{a_{2k}} (F_\beta^{-1} \:\Psi^{(h)}_{a_1\cdots 
a_{2k}}) \equiv 0$, i.e. $M_{a_{2k}}^{(\beta) \:\dagger} 
\psi^{(h)}_{a_1\cdots a_{2k}}\equiv 0$.
Analogously, the general solution of $Q_\beta \Psi=0$, i.e.
\begin{equation}
	(2k-1) \:M^{(\beta)}_{[a_1} \:\psi_{a_2 \cdots a_{2k-1}]} \;=\; 
	D_{a_{2k}} \:\psi_{a_1 \cdots a_{2k}} \;\;\;,
	\label{17}
\end{equation}
is given by
\begin{equation}
	\psi_{a_1 \cdots a_{2k-2}} \;=\; F_\beta^{-1} \:(N^\dagger
\:N)^{-1} 
	\:N_a^{(\beta) \:\dagger} \:D_b^{(\beta)} \:F_\beta\:\psi_{a_1
\cdots 
	a_{2k-2}a b}\:+\: \psi^{[h]}_{a_1 \cdots a_{2k-2}}
	\label{18}
\end{equation}
with $M_{[a}^{(\beta)} \:\psi_{a_1 \cdots a_{2k-2}]}^{[h]}\equiv0$.

Perhaps it is useful to present one of the proofs (e.g. that 
(\ref{15}) satisfies (\ref{12})) explicitly:
\begin{eqnarray}
\lefteqn{ N_{a_{2k}}^{(\beta)\:\dagger}\:\tilde{\psi}_{a_1 \cdots a_{2k}} 
\:-\: (2k-1) \:D_{[a_1}^{(\beta) \:\dagger} \:\tilde{\psi}_{a_2 \cdots 
a_{2k-1}]} } \label{19} \\
	 & = & -(2k-1) \:N_{a_{2k}}^{(\beta)\:\dagger} \:
N^{(\beta)}_{[a_{2k-1}} 
	 \: (N^\dagger \:N)^{-1}\: D_{(a_{2k})}^{(\beta)\:\dagger} \:
\tilde{\psi}_{a_1 
	 \cdots a_{2k-2}]} \nonumber \\
	 &&-(2k-1)(2k-2) \:(N^\dagger \:N)^{-1} \:N_{a_{2k}}^{(\beta)
\:\dagger} 
	 \:N^{(\beta)}_{[a_1} \: D^\dagger_{a_2} \:\tilde{\psi}_{a_3
\cdots ]
	 a_{2k}} \nonumber \\
	 &=& -(2k-1)  \:(N^\dagger \:N)^{-1} N^{(\beta)}_{[a_1} \;\left\{
	 \vec{N}^\dagger \:\vec{D}^\dagger \: \tilde{\psi}_{a_2 \cdots 
	 a_{2k-1}]} \:+\: (2k-2) \: [N^{(\beta) \:\dagger}_{(a_{2k})},
D^{(\beta) 
	 \:\dagger}_{a_2}] \: \tilde{\psi}_{a_3 \cdots ] a_{2k}}
	 \right. \nonumber \\
	 &&\left.
	 \:+\: (2k-2) \:D^{(\beta) \:\dagger}_{a_2}
\:N_{(a_{2k})}^{(\beta) \:\dagger} \: 
	 \tilde{\psi}_{a_3 \cdots ] a_{2k}} \right\} \;=\; 0 \;\;\;, 
	 \nonumber
\end{eqnarray}
as the first two terms inside the bracket combine to give $(-i z_E 
J_E \tilde{\psi})^{(2k-2)}$ (which is zero) and the last term vanishes 
by induction hypothesis (i.e. $\tilde{\psi}^{(2k-2)}$ satisfying 
$(\ref{12})_{k \rightarrow k-1}$).\\[.5cm]
{\em{Note added:}}
Instead of trying to give an analytical meaning to $N^{(\beta)}_{\alpha A}
(N^\dagger
N)^{-1}$, one may simply use
\[ I^{(\beta)}_{\alpha A} := \delta_{\alpha \beta} \: \frac{i q_A}{q^2}
\spc , \]
as $N_a^\dagger I_a = \1$ and $[I_a, N^\dagger_b]=0$ (as well as $[I_a,
F_\beta]=0$,
making unnecessary the detour via $F$).\\[.5em]
{\em{Acknowledgement:}} I am very grateful to F.~Finster, J.~Goldstone,
H. Nicolai, S.-T.~Yau and E.~Zaslow for helpful discussions and to the
Mathematics Department of Harvard University for its hospitality.

\end{document}